\documentclass[lettersize,journal]{IEEEtran}
\usepackage{amsmath,amsfonts}
\usepackage{algorithmic}
\usepackage{algorithm}
\usepackage{array}
\usepackage[caption=false,font=normalsize,labelfont=sf,textfont=sf]{subfig}
\usepackage{textcomp}
\usepackage{stfloats}
\usepackage{url}
\usepackage{verbatim}
\usepackage{graphicx}
\usepackage{cite}
\usepackage{multirow}
\usepackage{color}
\begin{document}

\title{Communication Delay Robust Control of BESS for AI Training Load Smoothing}

\author{Xue Lyu,~\IEEEmembership{Member,~IEEE,} Wei Du,~\IEEEmembership{Senior Member,~IEEE,} Sheik Mohammad Mohiuddin, ~\IEEEmembership{Member,~IEEE,} Brett A. Ross, ~\IEEEmembership{Member,~IEEE} 
\thanks{This work is supported by the U.S. Department of Energy's Office of Electricity under, "Foundational Studies and Technical Solutions for Large Digital Dynamic Loads." Pacific Northwest National Laboratory (PNNL) is operated by the Battelle for the U.S. Department of Energy under Contract DE-AC05-76RL01830. (Corresponding author: Wei Du.)

Xue Lyu, Sheik Mohammad Mohiuddin, and Brett A. Ross are with PNNL, Richland, WA 99352 USA (e-mail: xue.lyu@pnnl.gov;
sheik.mohiuddin@pnnl.gov; brett.ross@pnnl.gov).

Wei Du is with PNNL, Richland, WA 99352 USA, and also with Washington State University, Pullman, WA 99163
USA (e-mail: wei.du@pnnl.gov).
}
}

\markboth{Journal of \LaTeX\ Class Files,~Vol.~14, No.~8, August~2021}%
{Shell \MakeLowercase{\textit{et al.}}: A Sample Article Using IEEEtran.cls for IEEE Journals}


\maketitle

\begin{abstract}
AI training loads can exhibit rapid power fluctuations because their power demand differs significantly between computational and communication phases, creating challenging ramp rates at the data center point of common coupling (PCC). Integrating battery energy storage systems (BESS) in data centers is a promising mitigation option. This paper proposes a hybrid BESS control strategy that combines droop-based grid-forming (GFM) control with instantaneous load current-based compensation to suppress high frequency load fluctuations. 
The GFM control loop regulates the long-term power exchange of the BESS, while the load-following control provides fast compensation for short-term AI workload fluctuations. 
Communication delay between the load current measurements and the BESS controller is explicitly modeled, and its impact on smoothing performance is analyzed. To mitigate delay-induced degradation, a predictor-based compensation method is incorporated into the BESS control structure. High-fidelity electromagnetic transient simulations are conducted to validate the proposed approach. Results demonstrate effective smoothing of AI training load fluctuations across different grid strength conditions and under time varying communication delays.
\end{abstract}


\section{Introduction}
\IEEEPARstart{E}{merging} large loads have grown rapidly in recent years, with forecasts predicting that data centers alone may account for as much as 12\% of total US electricity consumption by 2028 \cite{force2025characteristics, shehabi20242024}. 
Their electrical behavior is less predictable than that of conventional loads, introducing new challenges for grid stability and reliability. 
A review of large artificial intelligence (AI) data center loads in \cite{seshmasetti2025review} identifies several key issues, including multi-timescale channel propagation, new oscillatory phenomena, voltage sensitivity, and protection miscoordination. 
Among the most pressing concerns are the rapid fluctuations in power consumption generated by AI training workloads. A single training job can span more than a hundred thousand Graphics Processing Units (GPUs)~\cite{choukse2025power}. During such a job, power demand surges in the computationally intensive phases and drops during the inter-node communication phases. At scale, these swings can amount to hundreds of megawatts, with transitions occurring on timescales of milliseconds to seconds. The existing grid was not designed to accommodate such load patterns, and new technologies are needed to address them. 

The operating characteristics of AI accelerators, such as GPUs, and their power consumption patterns are analyzed in \cite{li2025ai}. In \cite{ko2025wide}, stochastic and periodic power consumption profiles of AI workloads are characterized using a stochastic modeling framework, and simulation results show that severe oscillations can arise when data center-induced fluctuations excite unstable intrinsic modes of the grid. These results demonstrate the need for direct mitigation of data center load fluctuations. Several prior studies have also investigated the use of data center load flexibility to enhance frequency stability. In~\cite{10551465}, a multi-data center tie-line power smoothing method is proposed in which server clusters and uninterruptible power supply (UPS) units serve as demand response resources. In~\cite{8782131}, job scheduling policies and UPS batteries are coordinated through an optimization framework to provide grid frequency support. In~\cite{11353930}, delay tolerant (batch) workloads are treated as demand side flexibility in data centers and incorporated into a frequency secured unit commitment (UC) formulation. However, these approaches rely on workload scheduling or energy management strategies that operate on timescales of minutes to hours and are therefore unable to mitigate the sub-second to second power fluctuations generated by AI training loads. Google has incorporated instrumentation into the Tensor Processing Unit (TPU) compiler \cite{gan2025balance}, showing that, with proper tuning of the mitigation parameters, the magnitude of power fluctuations can be reduced by 47\%. Beyond software level methods, fast acting power electronics based solutions with real time control are also needed to effectively address this problem.

The batteries of a typical data center UPS have short storage duration, on the order of minutes. They are therefore not intended to provide sustained, repeated load ramp mitigation or long duration support during utility disturbances, for which dedicated battery energy storage system (BESS) installations offer greater energy capacity and operational flexibility.
In particular, a grid-forming (GFM) BESS appears to be a strong candidate because of its fast dynamic response. \cite{11225523} shows that integrating GFM BESS with AI-driven data centers can mitigate voltage sag and flicker caused by fluctuating loads.
In \cite{11407489}, a virtual synchronous machine (VSM)-based BESS was used to evaluate its effectiveness in smoothing AI training load fluctuations. 
Simulation results showed that the BESS could partially mitigate the load fluctuations, but full compensation remained difficult to achieve. A grid-forming BESS can compensate for only 40\%–60\% of the load fluctuations according to \cite{EPC_AI_DC_BESS}, depending on grid conditions. One BESS original equipment manufacturer (OEM) deployment shows a 70\% reduction in power variability when properly controlled \cite{ERCOT_Megapack_2025}. Supercapacitors provide another complementary solution due to their high power density and fast response time. However, their limited energy capacity makes it challenging for them to provide standalone smoothing for large scale data centers, particularly those with hundreds of megawatts or gigawatt scale demand.
Energy-storage STATCOMs (E-STATCOMs) provide another potential mitigation option, and a summary of available mitigation approaches is provided in \cite{samarawickrama2026esig}. Overall, these findings indicate that although BESSs are promising for mitigating AI training load fluctuations, the control design required to achieve high performance smoothing remains an open challenge.

The existing BESSs control approaches usually rely on plant level power commands that are sent to inverter level controllers through a communication network. This architecture introduces non-negligible end-to-end delay, which depends on telemetry bandwidth, network topology, routing, and the communication technologies and protocols used \cite{bhattarai2020studying}. A prior study \cite{fan2022analysis} has shown that the communication delay between the plant level control and the inverter level control is identified as one of the critical factors for the oscillations. A first-order Padé approximation is used in \cite{montero2022small} to model the communication delay, and it identifies that the interaction between active power control, computation delays, and inverter control potentially results in instability. Existing analyses commonly model communication delay as a fixed value. However, in practical data center applications, communication delay is typically time varying rather than constant.

Accounting for the above-mentioned challenges, this work proposes a new BESS control strategy to mitigate rapid power fluctuations caused by AI workloads. Unlike conventional plant level control approaches that rely on dispatch commands from a centralized plant controller to inverter level controllers, the proposed method uses instantaneous load current measurements to generate fast compensation references for the BESS inverter. This enables the BESS to respond to rapid load variations. Because the transmission of load current signals to the BESS controller is subject to communication latency, this work explicitly models the varying communication delay and develops a predictor-based compensation method to improve smoothing performance under delayed signal transmission. In addition, a GFM control layer is incorporated to regulate the long term active power output of the BESS and ensure stable operation under different grid strength conditions. The contributions of this work are twofold: (1) A hybrid BESS controller that combines GFM control with instantaneous load current-based compensation to suppress load variations is designed. The load-following control effectively damps the rapid power fluctuations characteristic of AI training loads on a short time scale, while the GFM controller regulates power output on a longer time scale and ensures reliable operation across different grid strength conditions. (2) The communication delays between the load current and the BESS controller are taken into account. A time varying communication delay is modelled, and a predictor-based compensation method is developed to mitigate its adverse effects and improve power smoothing performance. 

The rest of this paper is organized as follows. Section II presents the proposed hybrid control strategy, which combines droop-based GFM control with instantaneous load current compensation. Section III introduces the time varying communication delay model and the proposed predictive control design. Section IV evaluates the performance of the proposed control under different communication delay conditions, grid strengths, and AI training load profiles. Section V concludes the paper.
\vspace{-0.3cm}
\section{The Proposed Hybrid Control}
\subsection{System Model}
\begin{figure}
    \centering
    \includegraphics[width=0.5\textwidth]{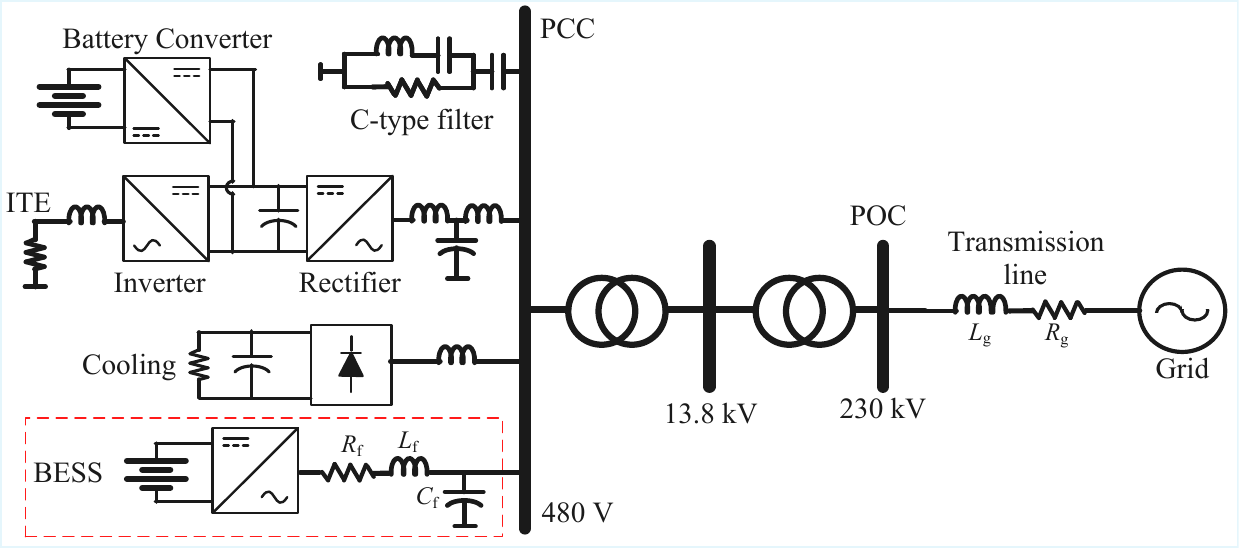}
    \vspace{-0.5cm}
    \caption{Schematic of the data center system connected to the transmission grid.}
    \vspace{-0.3cm}
    \label{fig:system_conf}
\end{figure}

The schematic of the data center system connected to the transmission grid is given in Fig.~\ref{fig:system_conf}. The data center model comprises cooling loads, information technology equipment (ITE), and a BESS. 
All components are connected at the 0.48\,kV point of common coupling (PCC). A C-type filter is installed at the PCC to mitigate harmonic distortion and provide reactive power compensation. The filter is designed to present low loss at the fundamental frequency while offering low impedance around targeted harmonic frequencies, thereby reducing harmonic current injection from nonlinear data center loads into the grid.
Two step-up transformers interface the PCC with a 230\,kV high voltage bus at the point of connection (POC). The transmission line impedance, characterized by~$R_g$ and~$X_g$, can be adjusted to represent strong or weak grid conditions. The transmission grid is represented by an infinite bus.

The cooling motor loads are simplified as resistive loads and integrated via a diode rectifier. The ITE is interfaced to the grid through a UPS, whose controls are modeled as follows:
\begin{itemize}
  \item \emph{Rectifier} - modeled as a grid-following (GFL) inverter with current vector control. The UPS rectifier operates at unity power factor and provides minimal grid support.
  \item \emph{Inverter} - modeled as a GFM inverter with droop-based control. A V/f control is utilized so that the inverter forms a regulated voltage bus for the ITE load.
  \item \emph{Battery converter} - modeled as a bidirectional buck–boost converter with two control loops: a DC voltage regulator and a power regulator.
\end{itemize}
The detailed control topology can be found in \cite{ross2026electromagnetic}.

\begin{figure*}
    \centering
    \includegraphics[width=0.85\textwidth]{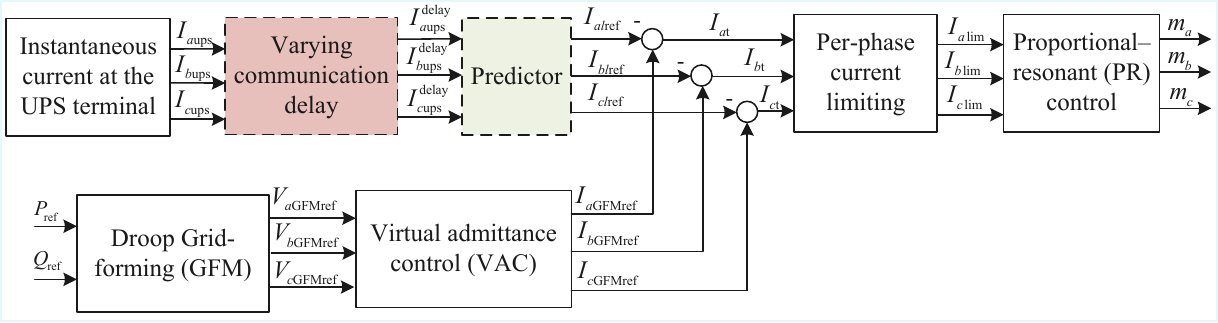}
    \vspace{-0.3cm}
    \caption{Overview of the proposed hybrid control structure.}
    \vspace{-0.3cm}
    \label{fig:control_overview}
\end{figure*}

A BESS is also installed at the data center to provide fast dynamic support to the grid. In this work, a hybrid control strategy is developed for the BESS to mitigate fluctuations caused by the AI training load. The overall structure of the proposed hybrid control scheme is illustrated in Fig.~\ref{fig:control_overview}. 
The GFM control determines the active and reactive power dispatch of the BESS on a long timescale, while the load-following-based control mitigates AI training load fluctuations on a short timescale. The detailed control design is presented in the following subsections. 
\vspace{-0.2cm}
\subsection{Hybrid GFM Control and Load Following Control}

\begin{figure}
    \centering
    \includegraphics[width=0.45\textwidth]{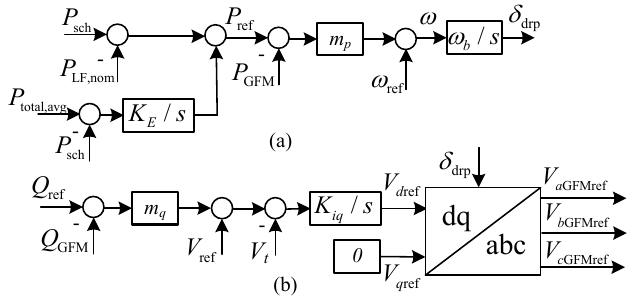}
    \vspace{-0.4cm}
    \caption{Droop grid-forming control, (a) phase angle, (b) voltage reference.}
    \vspace{-0.3cm}
    \label{fig:droop_GFM}
\end{figure}

The droop-based GFM control as given in Fig.~\ref{fig:droop_GFM} is employed in this work to establish the phase angle \(\delta_{\mathrm{drp}}\) and the voltage references of the BESS. The active power droop control determines the frequency as,
\begin{align}
    \dot{\delta}_{\mathrm{drp}} = \omega_{\mathrm{drp}} &= \omega_b + m_p\left(P_{\mathrm{ref}} - P_{\mathrm{GFM}}\right),
\end{align}
while the voltage control is given by,
\begin{align}
    \dot{V}_{d\mathrm{GFMref}} &= K_{iq}\left[m_q\left(Q_{\mathrm{ref}} - Q_{\mathrm{GFM}}\right) + V_{\mathrm{ref}} - V_t \right], \\
    V_{q\mathrm{GFMref}} &= 0.
\end{align}
Here, \(\omega_{\mathrm{drp}}\) is the angular frequency of the BESS inverter, \(\omega_b\) is the nominal grid angular frequency, and \(V_t\) is the measured magnitude of the LC filter terminal voltage. \(V_{d\mathrm{GFMref}}\) and \(V_{q\mathrm{GFMref}}\) are the \(d\)-axis and \(q\)-axis voltage references, respectively, and \(K_{iq}\) is the integral gain of the voltage control loop. The parameters \(m_p\) and \(m_q\) denote the droop coefficients for active and reactive power control, respectively. 
\(P_{\mathrm{GFM}}\) and \(Q_{\mathrm{GFM}}\) are the active and reactive powers associated with the GFM control branch. Using \(P_{\mathrm{GFM}}\) as the droop feedback decouples the GFM power regulation loop from the fast load current compensation loop.

In this work, the active power reference of the GFM branch $P_{\mathrm{ref}}$ is generated by a slow supervisory loop as,
\begin{align}
    \dot{\xi}_E &= P_{\mathrm{total,avg}} - P_{\mathrm{sch}}, \\
    P_{\mathrm{ref}} &=
    P_{\mathrm{sch}}
    - P_{\mathrm{LF,nom}}
    - K_E \xi_E ,
\end{align}
where \(P_{\mathrm{sch}}\) is the scheduled long-term total BESS power, \(P_{\mathrm{L,nom}}\) is the nominal power demand from the AI workload.
\(P_{\mathrm{total,avg}}\) is the low-pass-filtered total BESS output power with a large time constant \(T_{\mathrm{avg}}\), and \(K_E\) is the integral gain of the supervisory loop. Since the total BESS active power is composed of the GFM-branch power and the load-following compensation power, $P_{\mathrm{total}} = P_{\mathrm{GFM}} + P_{\mathrm{LF}},$
the term \(P_{\mathrm{LF,nom}}\) is subtracted from the scheduled total BESS power to determine the DC component (steady-state) of the BESS output power for the GFM branch. The integral term slowly corrects any mismatch between the scheduled total BESS power and the actual averaged BESS output power. A small \(K_E\) is selected so that this supervisory loop operates on a much slower time scale than the fast load-following compensation loop and therefore does not interfere with rapid power smoothing actions. 
In this way, the slow supervisory loop regulates the long term average power exchange of the BESS and prevents energy drift caused by the fast load-following compensation.

The three-phase voltage references in $abc$-frame $V_{a\mathrm{GFMref}}$, $V_{b\mathrm{GFMref}}$, and $V_{c\mathrm{GFMref}}$ are obtained from $V_{d\mathrm{GFMref}}$ and $V_{q\mathrm{GFMref}}$ through the inverse Park transformation using the angle $\delta_{\mathrm{drp}}$.


A virtual admittance control (VAC) \cite{taul2019current, zhang2023current} is utilized in this work to generate the three-phase current references $I_{a\text{GFMref}}$, $I_{b\text{GFMref}}$, and $I_{c\text{GFMref}}$ for the GFM control, as given by,
\begin{align}
    I_{a\text{GFMref}} & = \frac{V_{a\text{GFMref}} - V_{at}}{R_v + sL_v}, \\
    I_{b\text{GFMref}} & = \frac{V_{b\text{GFMref}} - V_{bt}}{R_v + sL_v}, \\
    I_{c\text{GFMref}} & = \frac{V_{c\text{GFMref}} - V_{ct}}{R_v + sL_v},
\end{align}
where $V_{at}$, $V_{bt}$, and $V_{ct}$ are the BESS  LC filter terminal voltage in the $abc$ frame, and $R_v$ and $L_v$ are the virtual resistance and virtual inductance, respectively. This VAC method could inherently emulate a virtual impedance between the voltage references generated by the droop GFM control and the terminal voltage.

As given in Fig.~\ref{fig:system_conf}, the AI training load is connected to the PCC through a centralized UPS, a topology commonly adopted by data center developers. A meter can be installed at the UPS terminal to measure the instantaneous load currents $I_{a\text{ups}}$, $I_{b\text{ups}}$, and $I_{c\text{ups}}$, which are then transmitted to the BESS controller. By using the negative of the measured instantaneous load current as a reference signal, the BESS injects a corresponding compensating current, thereby mitigating AI workload induced power fluctuations at the data center PCC. 
In addition to short-term fast fluctuation mitigation, long-term regulation of the BESS power output is also desired. To achieve both objectives, the instantaneous load current compensation reference and the current reference from the GFM controller are combined to generate the total BESS current references $I_{at}$, $I_{bt}$, and $I_{ct}$, as shown in Fig.~\ref{fig:control_overview}.

\subsection{Per-Phase Current Limiting}
\begin{figure}
    \centering
    \includegraphics[width=0.5\textwidth]{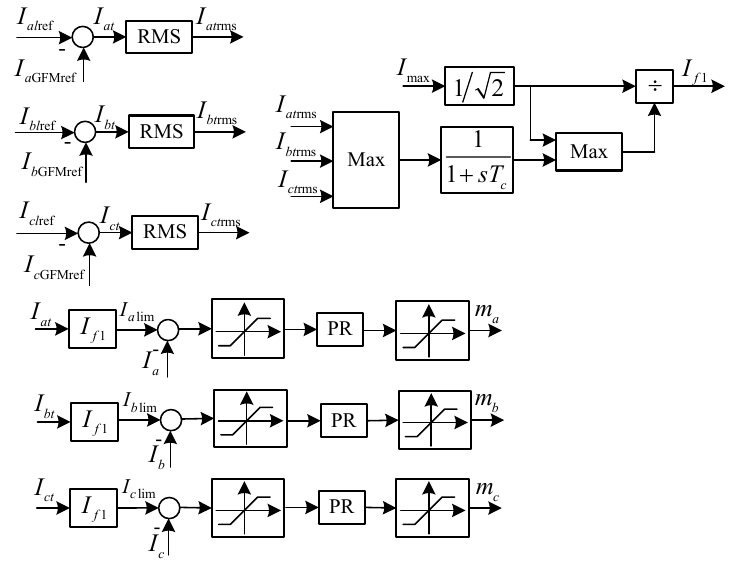}
    \vspace{-0.5cm}
    \caption{Per-phase current limiting and modulation index generation.}
    \vspace{-0.3cm}
    \label{fig:per_phase_CL}
\end{figure}

To ensure that the current injected by the BESS remains within its operating limits, a current limiting mechanism is required. In this work, a per-phase current limiting method as shown in Fig.~\ref{fig:per_phase_CL} is employed. The Root Mean Square (RMS) values of the total three-phase current references, $I_{at\mathrm{rms}}$, $I_{bt\mathrm{rms}}$, and $I_{ct\mathrm{rms}}$, are continuously monitored. When the RMS current in any phase exceeds the maximum allowable value, each phase current reference is scaled proportionally according to,
\begin{align}
I_{a\text{lim}} &= I_{f1} I_{at},\\
I_{b\text{lim}} &= I_{f1} I_{bt}, \\
I_{c\text{lim}} &= I_{f1} I_{ct}, \\
I_{f1} &=
\frac{I_{\max}/\sqrt{2}}
{\max \left( I_{\max}/\sqrt{2},\  \frac{1}{1+sT_c}\max \left[ I_{at\mathrm{rms}},\, I_{bt\mathrm{rms}},\, I_{ct\mathrm{rms}} \right] \right)},
\end{align}
where $I_{a\text{lim}}$, $I_{b\text{lim}}$, and $I_{c\text{lim}}$ are the limited current references for phases $a$, $b$, and $c$, respectively, $I_{\max}$ is the maximum allowable peak current, and $T_c$ is the time constant of the low-pass filter. The scaling factor $I_{f1}$ is applied equally to all three phases so that the relative phase relationships among the current references are preserved while limiting the overall current magnitude. The low-pass filter is introduced to avoid abrupt changes in the scaling factor and to improve the smoothness of the limiter response. In this way, the proposed current limiting method ensures that the BESS current injection remains within its allowable operating range.

The output of the per-phase current limiter is fed to a proportional-resonant (PR) controller, from which the PWM modulation indices $m_a$, $m_b$, and $m_c$ are generated,
\begin{align}
m_{a} = \left[K_p + \frac{2K_i \omega_{\text{cut}} s}{s^2 + 2\omega_{\text{cut}} s + \omega_b^2}\right]\left(I_{a\text{lim}} - I_{a}\right), \\
m_{b} = \left[K_p + \frac{2K_i \omega_{\text{cut}} s}{s^2 + 2\omega_{\text{cut}} s + \omega_b^2}\right]\left(I_{b\text{lim}} - I_{b}\right),\\
m_{c} = \left[K_p + \frac{2K_i \omega_{\text{cut}} s}{s^2 + 2\omega_{\text{cut}} s + \omega_b^2}\right]\left(I_{c\text{lim}} - I_{c}\right),
\end{align}
where $I_a$, $I_b$, and $I_c$ are the actual BESS output currents in phases $a$, $b$, and $c$, respectively; $K_p, K_i$ are gain constants of the PR controller, and $\omega_{\text{cut}}$ is cutoff frequency. 

\vspace{-0.2cm}
\section{Communication Delay Modeling and Predictor Control Design}
This section presents the communication delay model associated with transmitting the load current signal to the BESS controller. In particular, stochastic varying communication delay is considered in this work. To mitigate the degradation in power smoothing performance caused by this delay in the load following control, a predictor-based compensation strategy is developed.

\subsection{Time Varying Communication Delay Model}
The end-to-end communication latency consists of several components, including propagation delay, transmission delay, queuing delay, and processing delay \cite{DOE_SPARC_Latency_2024}. These components are influenced by multiple factors, such as the communication protocol, physical medium and access scheme, network size and architecture, traffic loading, and the performance and configuration of communication devices. Low communication latency is critical for real-time monitoring, control, and coordination of grid assets, particularly in applications requiring fast response. For the data center application considered in this work, high-speed communication is therefore essential. 

To capture realistic communication behavior, the communication delay is modeled as the sum of a fixed component and a varying component. The fixed component represents the fixed system latency, while the varying component captures delay variations caused primarily by transmission and network-induced effects. Accordingly, the communication delay is expressed as,
\begin{equation}
T_{dv} = T_0 + \Delta T_d,
\label{eq:delay_model}
\end{equation}
where $T_0$ is the fixed delay and $\Delta T_d \in [\Delta T_{\min}, \Delta T_{\max}]$ is the stochastic delay variation. In this work, $\Delta T_d$ is implemented as a bounded Gaussian random variable, generated by truncating a Gaussian distribution at a specified number of standard deviations and scaling it to lie within the prescribed minimum and maximum limits.

\begin{figure}
    \centering
    \includegraphics[width=0.2\textwidth]{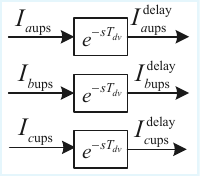}
    \vspace{-0.5cm}
    \caption{Time varying communication delay applied to the measured load current signals.}
    \vspace{-0.3cm}
    \label{fig:varying_delay}
\end{figure}

After the time varying communication delay, the BESS receives the delayed instantaneous load current signals $I_{a\text{ups}}^{\text{delay}}$, $I_{b\text{ups}}^{\text{delay}}$,  and $I_{c\text{ups}}^{\text{delay}}$ as shown in Fig.~\ref{fig:varying_delay}.
\subsection{Designed Predictive Control}
\begin{figure*}
    \centering
    \includegraphics[width=1\textwidth]{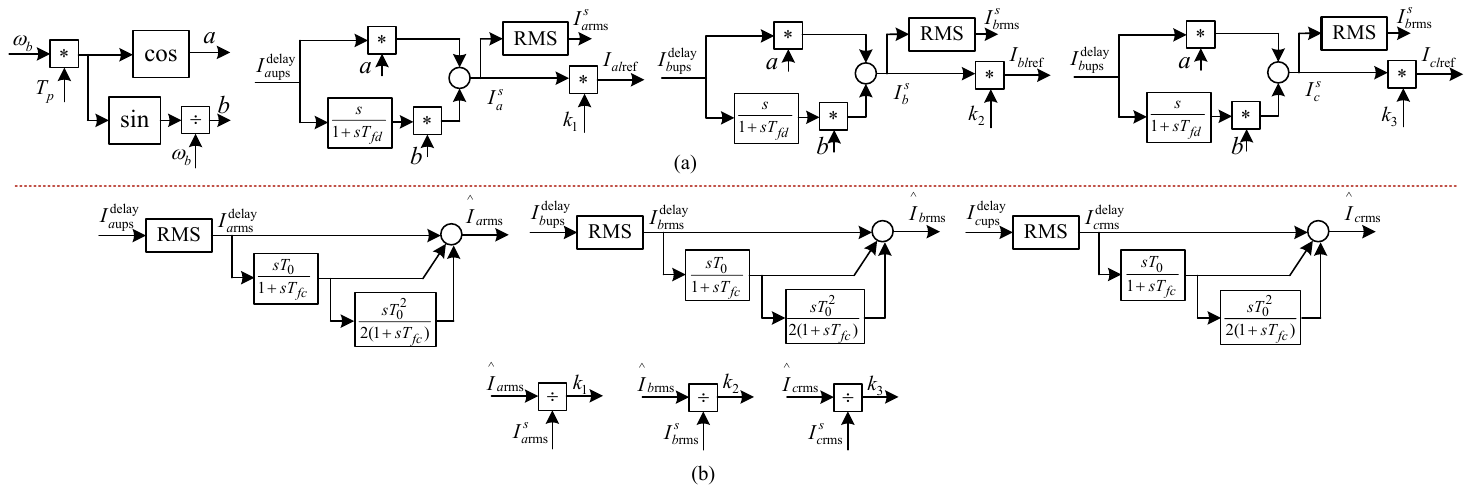}
    \vspace{-0.5cm}
    \caption{Structure of the proposed predictor-based delay compensation scheme, (a) phase compensation, (b) magnitude compensation.}
    \vspace{-0.3cm}
    \label{fig:predictive_control}
\end{figure*}

Communication delay introduces phase lag into the measured load current signal and degrades the power smoothing performance of the BESS. 
To mitigate this effect, a predictor-based compensation method is developed, as shown in Fig.~\ref{fig:predictive_control}. The proposed predictive control consists of two parts: 1) a sinusoidal predictor that compensates for the phase lag of the instantaneous phase currents, and 2) an RMS-current predictor that corrects the load current magnitude to improve tracking during rapid load changes.

The instantaneous load current is dominated by the grid fundamental frequency component. Therefore, for each phase, the delayed current can be advanced to estimate the present instantaneous current. Using a sinusoidal predictor, the predicted current in each phase is expressed as,
\begin{align}
    I_a^s &= I_{a\mathrm{ups}}^{\mathrm{delay}} \cos(\omega_b T_p)
    + \frac{1}{\omega_b}\frac{s}{1+sT_{fd}}I_{a\mathrm{ups}}^{\mathrm{delay}} \sin(\omega_b T_p), \\
    I_b^s &= I_{b\mathrm{ups}}^{\mathrm{delay}} \cos(\omega_b T_p)
    + \frac{1}{\omega_b}\frac{s}{1+sT_{fd}}I_{b\mathrm{ups}}^{\mathrm{delay}} \sin(\omega_b T_p), \\
    I_c^s &= I_{c\mathrm{ups}}^{\mathrm{delay}} \cos(\omega_b T_p)
    + \frac{1}{\omega_b}\frac{s}{1+sT_{fd}}I_{c\mathrm{ups}}^{\mathrm{delay}} \sin(\omega_b T_p),
\end{align}
where \(\omega_b\) is the nominal grid angular frequency, \(T_{fd}\) is the time constant of the filtered derivative used to reduce noise amplification, and \(T_p\) is the delay input used in the predictive control. 

The sinusoidal predictor is equivalent to a phase advance operation for the fundamental component. For a delayed sinusoidal current \(i^{\mathrm{delay}}(t)=I_m\cos(\omega_b t+\phi-\omega_b T_d)\), where \(I_m\) and \(\phi\) are the current magnitude and phase angle, respectively, the ideal predictor gives,
\begin{align}
    i^s(t)
    &= i^{\mathrm{delay}}(t)\cos(\omega_b T_p)
    +\frac{1}{\omega_b}\frac{d i^{\mathrm{delay}}(t)}{dt}\sin(\omega_b T_p) \nonumber \\
    &= I_m\cos(\omega_b t+\phi-\omega_b T_d)\cos(\omega_b T_p) \nonumber \\
    &\quad - I_m\sin(\omega_b t+\phi-\omega_b T_d)\sin(\omega_b T_p) \nonumber \\
    &= I_m\cos\left(\omega_b t+\phi-\omega_b T_d+\omega_b T_p\right).
\end{align}
Therefore, the predictor advances the delayed signal by \(T_p\). When \(T_p=T_d\), \(i^s(t)=i(t)\), indicating exact delay compensation for an ideal fundamental frequency sinusoid. When \(T_p\neq T_d\), the residual phase error is \(\omega_b(T_p-T_d)\). 
In this work, \(T_p\) is determined by sweeping the predictor delay within the admissible interval as shown in \eqref{eq:delay_model} and choosing the value that yields the best overall smoothing performance in terms of broadband fluctuation reduction and ramp rate suppression. This approach is particularly useful when the communication delay varies within a known range but is not measured online. In such cases, selecting \(T_p\) based on performance tuning improves robustness under varying delay conditions.

In practice, the derivative is implemented using a filtered derivative $\frac{s}{1+sT_{fd}}$ to avoid excessive noise amplification. Consequently, the compensation is not mathematically exact because the filtered derivative introduces magnitude attenuation and phase lag. For the fundamental frequency component, the filtered derivative has the frequency response
\begin{equation}
    D_f(j\omega_b)
    =
    \frac{j\omega_b}{1+j\omega_b T_{fd}}.
\end{equation}
When \(\omega_b T_{fd}\ll 1\), this response approaches the ideal derivative \(j\omega_b\), and the sinusoidal predictor provides an accurate phase-advance approximation. Therefore, the predictor is exact only for an ideal sinusoidal current and an ideal derivative, but it remains effective for practical currents whose dominant component is the grid fundamental, provided that \(T_{fd}\) is properly selected.

Although the sinusoidal predictor effectively compensates for the phase lag of the fundamental frequency waveform, it may not fully capture the current magnitude variation caused by sharp load transitions. This limitation is important for AI workload profiles, which can exhibit abrupt changes over short time intervals. To improve current magnitude tracking during such transitions, the RMS value of each delayed phase current is measured and extrapolated forward.
Since the RMS current is an averaged quantity, a Taylor-type second-order extrapolation with filtered derivative terms is adopted,
\begin{align}
    \hat{I}_{a\mathrm{rms}}
    &= I_{a\mathrm{rms}}^{\mathrm{delay}}
    + \frac{sT_r}{1+sT_{fc}} I_{a\mathrm{rms}}^{\mathrm{delay}}
    + \frac{s^2T_r^2}{2(1+sT_{fc})} I_{a\mathrm{rms}}^{\mathrm{delay}}, \\
    \hat{I}_{b\mathrm{rms}}
    &= I_{b\mathrm{rms}}^{\mathrm{delay}}
    + \frac{sT_r}{1+sT_{fc}} I_{b\mathrm{rms}}^{\mathrm{delay}}
    + \frac{s^2T_r^2}{2(1+sT_{fc})} I_{b\mathrm{rms}}^{\mathrm{delay}}, \\
    \hat{I}_{c\mathrm{rms}}
    &= I_{c\mathrm{rms}}^{\mathrm{delay}}
    + \frac{sT_r}{1+sT_{fc}} I_{c\mathrm{rms}}^{\mathrm{delay}}
    + \frac{s^2T_r^2}{2(1+sT_{fc})} I_{c\mathrm{rms}}^{\mathrm{delay}},
\end{align}
where \(T_{fc}\) is the time constant used in the filtered extrapolation of the RMS current, and $T_r$ is the extrapolation horizon for the RMS-current predictor. In this work, $T_r$ is selected based on the fixed communication delay $T_0$.

The RMS values of the currents after the sinusoidal predictor, denoted by \(I_{a\mathrm{rms}}^s\), \(I_{b\mathrm{rms}}^s\), and \(I_{c\mathrm{rms}}^s\), are then measured. Finally, the current references are obtained by scaling the sinusoidally predicted currents according to the ratio between the extrapolated RMS current and the RMS current of the sinusoidal predictor,
\begin{align}
    I_{al\mathrm{ref}} = \frac{\hat{I}_{a\mathrm{rms}}}{I_{a\mathrm{rms}}^s} I_a^s, \quad I_{bl\mathrm{ref}} = \frac{\hat{I}_{b\mathrm{rms}}}{I_{b\mathrm{rms}}^s} I_b^s,  \quad I_{cl\mathrm{ref}} = \frac{\hat{I}_{c\mathrm{rms}}}{I_{c\mathrm{rms}}^s} I_c^s.
\end{align}

In this way, the sinusoidal predictor compensates for the phase lag of the delayed current waveform, while the RMS-current extrapolation corrects the current magnitude. The sinusoidal predictor mainly improves phase alignment, whereas the RMS-current predictor improves amplitude tracking during rapid load changes. Their combination provides a more accurate current reference for the BESS, thereby enhancing the power smoothing performance under time varying communication delay.
\vspace{-0.2cm}
\section{Case Studies}
The fixed parameters of the BESS controller are given in Table~\ref{Tab: parameter}. 
The proposed control strategy is evaluated through high-fidelity EMT simulations under different communication delays, grid strengths, and AI workload profiles. The results are presented and analyzed in this section.
\begin{table}
\caption{Parameters of the BESS inverter.}
\vspace{-0.2cm}
\centering
\begin{tabular}{ c|c||c|c  }
 \hline
 Parameter & Value & Parameter & Value\\
 \hline
 $m_p$  &  0.01  &  $m_q$ & 0.02 \\
 $R_c$ & 0.03 pu  & $L_c$ & 0.08 pu \\
 $R_v$ & 0.05 pu   & $X_v$  & 0.2 pu\\
 $I_{\text{max}}$ & 1 pu & $K_{iq}$ & 0.31 \\
 $K_p$   & 100 & $K_i$ &  5000 \\
 $\omega_b$ &  376.99 rad/s   & $\omega_{\text{cut}}$ & 1.57 rad/s \\
 $T_{fd}$& 0.001 s &  $T_c$ & 0.002 s \\
 $T_{\text{avg}}$ & 5 s  & $K_E$ & 0.01 \\
 \hline
\end{tabular} \label{Tab: parameter}
\end{table}
\subsection{Performance under Different Communication Delays}
According to a major BESS OEM in the U.S., the communication delay associated with transmitting load current signals to the BESS inverter can be maintained within 10 ms. To evaluate the impact of this delay, three scenarios are first compared: (1) the baseline case without BESS; (2) a 300 MW BESS equipped with the proposed control, assuming no communication delay in transmitting the load current signal to the BESS; and (3) a 300 MW BESS equipped with the proposed control, with a fixed 10 ms delay plus a stochastic delay variation of ± 1 ms, but without the proposed predictive compensation. 
All scenarios are simulated under a strong grid condition with the short circuit ratio (SCR) set to 10. A representative AI training load profile from \cite{choukse2025power} is used to evaluate the proposed control under these scenarios. 

\begin{figure}
    \centering
    \includegraphics[width=0.5\textwidth]{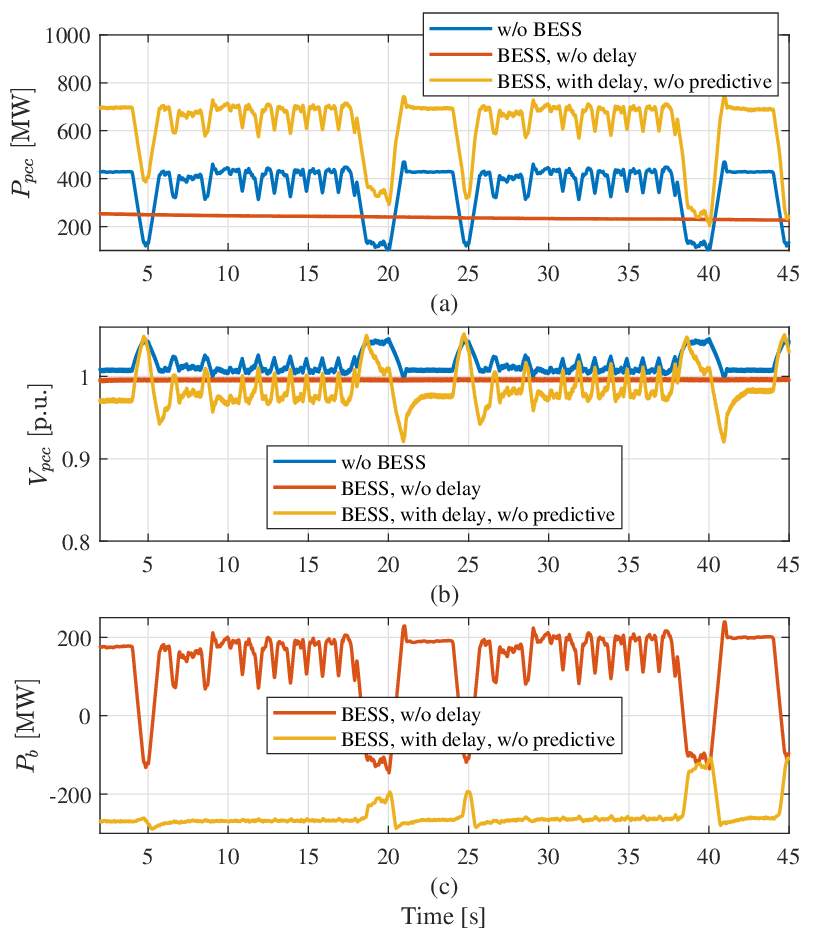}
    \vspace{-0.5cm}
    \caption{Simulation results comparing the baseline case without BESS, the case with BESS and no communication delay, and the case with BESS under 9 ms-11 ms communication delay without predictive control: (a) PCC active power, (b) PCC voltage magnitude, and (c) BESS active power.}
    \vspace{-0.3cm}
    \label{fig:ERCOT_3cases}
\end{figure}

The data center PCC active power, PCC voltage magnitude, and BESS active power for these scenarios are shown in Fig.~\ref{fig:ERCOT_3cases}. Compared with the baseline case without BESS, the proposed control fully compensates for the PCC power fluctuations under the ideal scenario with no communication delay. As a result, the voltage fluctuations at the data center PCC are effectively eliminated because the active power fluctuations are smoothed.
In contrast, when the communication delay varies between 9 ms and 11 ms, the proposed control without predictive compensation is unable to effectively smooth the AI training load fluctuations. 
This is because the communication delay introduces a significant phase lag in the transmitted load current signal. For example, a 10 ms delay corresponds to a phase lag of approximately $216^{\circ}$ at 60 Hz. 
This large phase lag can cause the BESS to inject a compensation current that is not properly aligned with the actual load current fluctuation, thereby degrading or even compromising the smoothing performance.

\begin{figure}
    \centering
    \includegraphics[width=0.5\textwidth]{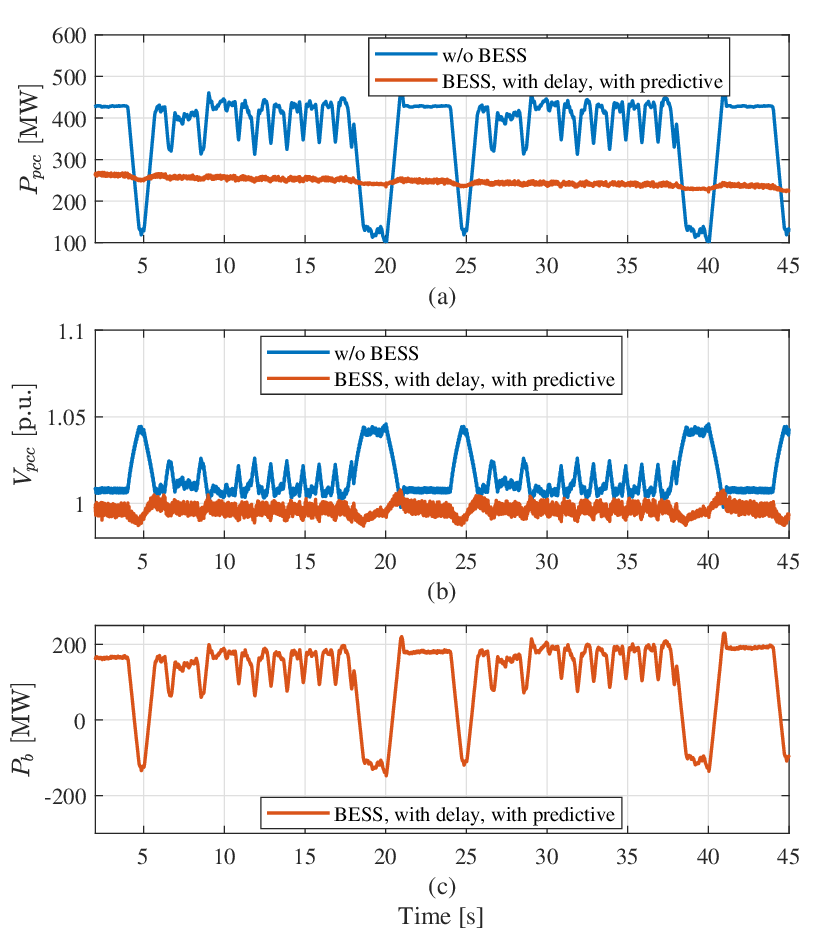}
    \vspace{-0.5cm}
    \caption{Simulation results for the BESS with the proposed predictive control under 9 ms-11 ms communication delay: (a) PCC active power, (b) PCC voltage magnitude, and (c) BESS active power.}
    \vspace{-0.5cm}
    \label{fig:ERCOT_delay10}
\end{figure}

\begin{figure}
    \centering
    \includegraphics[width=0.5\textwidth]{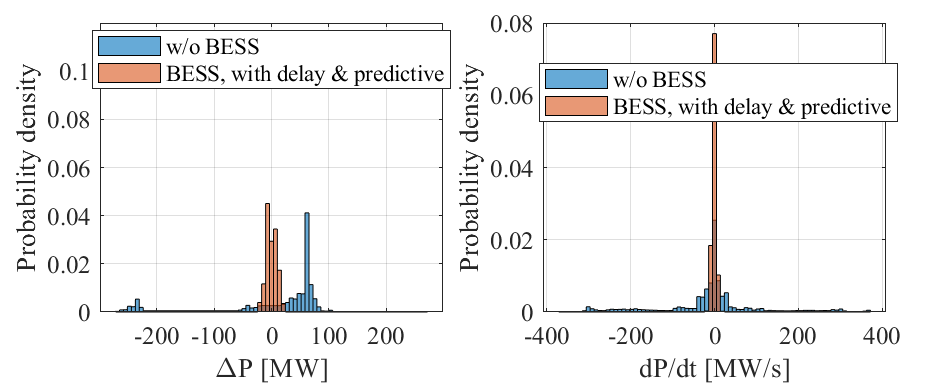}
    \vspace{-0.5cm}
    \caption{Distribution of the PCC active power deviations and ramp rates with and without BESS.}
    \vspace{-0.55cm}
    \label{fig:ERCOT_delay10_excur}
\end{figure}

\begin{figure}
    \centering
    \includegraphics[width=0.45\textwidth]{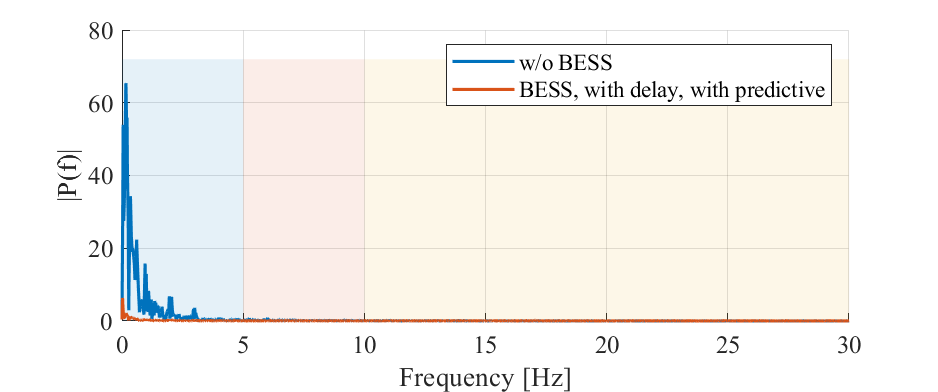}
    \vspace{-0.4cm}
    \caption{Single-sided amplitude spectrum of the PCC active power with and without BESS after DC component removal.}
    \vspace{-0.5cm}
    \label{fig:FFT_ERCOT_delay10}
\end{figure}

With the communication delay varying between 9 ms and 11 ms, the proposed predictive control as given in Fig.~\ref{fig:predictive_control} is applied with \(T_p = 10~\mathrm{ms}\) and \(T_{fc}=0.0016~\mathrm{s}\). The corresponding simulation results are shown in Fig.~\ref{fig:ERCOT_delay10}. It can be observed that the active power fluctuations at the data center PCC are significantly reduced with the BESS under the proposed predictive control. The power deviations and ramp rates are evaluated using a moving-average method with a 1 s time window. The resulting distributions of the power deviation and ramp rate are shown in Fig.~\ref{fig:ERCOT_delay10_excur}, and the corresponding quantitative comparison is summarized in Table~\ref{Tab:ramp_rate_ERCOT_delay10}. The single-sided amplitude spectrum of the PCC active power is calculated using a fast Fourier transform (FFT) with a Hann window. The DC component is removed before applying the FFT so that the resulting spectrum represents only the power-fluctuation content. The resulting spectrum is shown in Fig.~\ref{fig:FFT_ERCOT_delay10}, and the band-by-band reduction in the RMS fluctuation amplitude is summarized in Table~\ref{Tab:band_analysis_ERCOT_10}. Without the BESS, the peak-to-peak power excursion at the PCC reaches 374.82~MW, and the maximum ramp rate is 372.1~MW/s. With the BESS operating under the proposed predictive control, the peak-to-peak excursion is reduced by 87.5\%, and the 99th-percentile excursion is reduced by 91.9\%. Likewise, the maximum ramp rate is reduced by 94.4\%, and the 99th-percentile ramp rate by 96.0\%. 
Without the BESS, 97.84\% of the active-power fluctuation energy at the data-center PCC is concentrated in the 0.1-5~Hz range, and an additional 1.78\% lies within 5-10~Hz, indicating that these two bands dominate the fluctuation spectrum in this scenario. The proposed predictive control effectively attenuates the fluctuations in both bands, yielding an overall broadband fluctuation reduction of 95.2\% over the full simulation period. These results confirm that the proposed predictive control remains effective in smoothing AI training load fluctuations under time varying communication delay.
\begin{table}
\vspace{-0.2cm}
\caption{Comparison of PCC active power deviations and ramp rates.}
\vspace{-0.2cm}
\centering
\begin{tabular}{ c|c|c|c  }
 \hline
Metric & w/o BESS & with BESS & reduction\\
 \hline
Peak-to-peak excursion  & 374.82 MW   & 46.93 MW  & 87.5\%\\
 99th-pct-based deviation  & 253.82MW   & 20.61 MW  & 91.9\%\\
 Max $|dP/dt|$  & 372.1 MW/s & 21 MW/s & 94.4\% \\
 99th-pct $|dP/dt|$  & 330.5 MW/s & 13.3 MW/s &  96.0\% \\
 \hline
\end{tabular} \label{Tab:ramp_rate_ERCOT_delay10}
\vspace{-0.2cm}
\end{table}
\begin{table}
\vspace{-0.2cm}
\caption{Band-by-band analysis of PCC active power fluctuation reduction.}
\vspace{-0.2cm}
\footnotesize
\setlength{\tabcolsep}{4.5pt}
\centering
\begin{tabular}{ c|c|c||c|c|c }
 \hline
\shortstack {Frequency \\ band} & Weight & Reduction & \shortstack {Frequency \\ band} & Weight & Reduction\\
 \hline
 0.1-5 Hz  & 97.84\%   & 95.3\% & 5-10 Hz  & 1.78\% &  67.8\% \\
 10-30 Hz  & 3.6\% &  -23.9\% & 30-60 Hz  & 0.02\%  & -875.0\%\\
 \hline
\end{tabular} \label{Tab:band_analysis_ERCOT_10}
\end{table}

\begin{figure}
    \centering
    \includegraphics[width=0.5\textwidth]{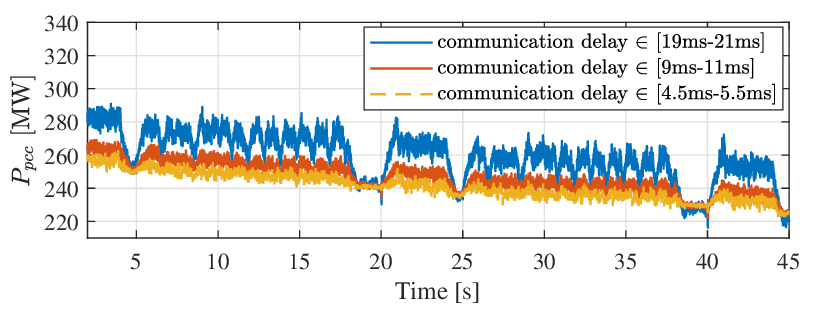}
    \vspace{-0.5cm}
    \caption{Data center PCC active power for the BESS with the proposed predictive control under different communication delay ranges.}
    \vspace{-0.3cm}
    \label{fig:different delay}
\end{figure}

Under \(\mathrm{SCR}=10\), different communication delay ranges are considered to evaluate the performance of the proposed predictive control. In addition to the communication delay varies between 9 ms and 11 ms, Fig.~\ref{fig:different delay} compares another two cases: (1) a delay varying between 4.5 ms and 5.5 ms, represented by a 5 ms fixed delay with a stochastic variation of \(\pm 0.5\) ms; and (2) a delay varying between 19 ms and 21 ms, represented by a 20 ms fixed delay with a stochastic variation of \(\pm 1\) ms. For the 4.5-5.5 ms delay case, the predictor delay is set to \(T_p=5.5\) ms and \(T_{fc}=0.0016\) s. For the 19-21 ms delay case, the predictor delay is set to \(T_p=20\) ms and \(T_{fc}=0.005\) s. The band-by-band analysis of AI workload fluctuation reduction is summarized in Table~\ref{Tab:band_analysis_ERCOT_different_delay}.The proposed control achieves a weighted broadband reduction of 96.1\% when the communication delay varies between 4.5 ms and 5.5 ms, and 89.8\% when the delay varies between 19 ms and 21 ms. The active power deviation and ramp rate comparisons are provided in Table~\ref{Tab:ramp_rate_ERCOT_different_delay}. These results show that the proposed predictive control significantly reduces both power deviations and ramp rates under different communication delay conditions. As expected, the smoothing performance is better under the lower delay condition, while the proposed control still maintains effective fluctuation mitigation capability even when the communication delay increases to the 19-21 ms range.
\begin{table}
\vspace{-0.2cm}
\caption{Band-by-band analysis of PCC active power fluctuation reduction with different communication delay ranges.}
\vspace{-0.2cm}
\footnotesize
\setlength{\tabcolsep}{4.5pt}
\centering
\begin{tabular}{ c|c|c}
 \hline
\shortstack {Frequency \\ band} & \shortstack{Reduction \\ delay $\in$ [4.5ms, 5.5ms] } & \shortstack{Reduction \\ delay $\in$ [19ms, 21ms]  } \\
 \hline
 0.1-5 Hz  & 96.1\%   & 89.9\%  \\
 5-10 Hz   & 59.8\%   & 41.3\%  \\
 10-30 Hz  & -23.9\%   & -103.6\%  \\
 30-60 Hz  & -690.9\% & -1086.5\% \\
 Weighted broadband & 95.9\%  & 89.8\%  \\
 \hline
\end{tabular} \label{Tab:band_analysis_ERCOT_different_delay}
\end{table}
\begin{table}
\vspace{-0.2cm}
\caption{Comparison of PCC active power deviations and ramp rates with different communication delay ranges.}
\vspace{-0.2cm}
\footnotesize
\setlength{\tabcolsep}{4.5pt}
\centering
\begin{tabular}{ c|c|c  }
 \hline
Metric  & \shortstack{Reduction \\ delay $\in$ [4.5, 5.5ms] } & \shortstack{Reduction \\ delay $\in$ [19, 21ms]  }\\
 \hline
Peak-to-peak excursion   & 89.2\%  & 80.1\%\\
99th-pct-based deviation & 93.1\%  & 85.8\%\\
Max $|dP/dt|$   & 96.2\% & 86.7\% \\
99th-pct $|dP/dt|$  & 97.2\% & 89.6\% \\
 \hline
\end{tabular} \label{Tab:ramp_rate_ERCOT_different_delay}
\end{table}
\vspace{-0.2cm}
\subsection{Performance under Different Grid Strengths}
\begin{figure}
    \centering
    \includegraphics[width=0.5\textwidth]{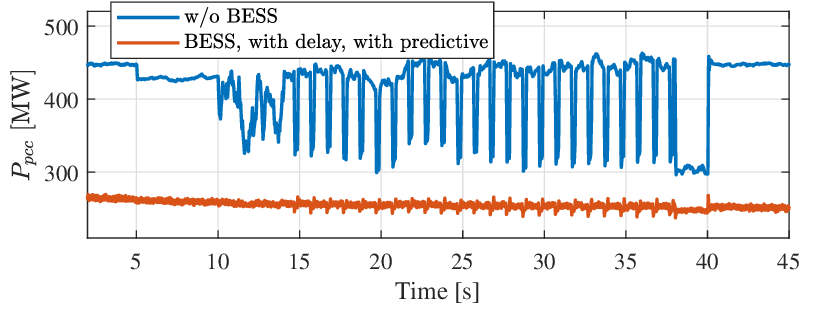}
    \vspace{-0.5cm}
    \caption{Data center PCC active power with and without the BESS for the NREC AI training load profile.}
    \vspace{-0.3cm}
    \label{fig:NREC_load}
\end{figure}

\begin{figure}
    \centering
    \includegraphics[width=0.5\textwidth]{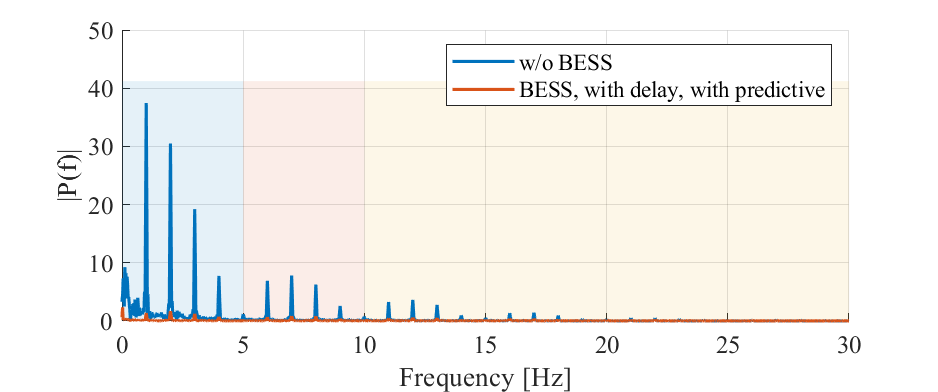}
    \vspace{-0.45cm}
    \caption{Single-sided amplitude spectrum of the PCC active power after DC component removal for the NREC AI training load profile.}
    \vspace{-0.3cm}
    \label{fig:Nrecload_FFT}
\end{figure}

An example AI data center load profile from the NREC Large Load Task Force white paper~\cite{force2025characteristics} is used to test the proposed control.
Fig.~\ref{fig:NREC_load} shows the data center PCC active power with and without the 300 MW BESS. The simulation is conducted under \(\mathrm{SCR}=3\), with the communication delay varying between 9 ms and 11 ms.
The single-sided amplitude spectrum of the PCC active power after DC component removal, obtained using FFT analysis, is shown in Fig.~\ref{fig:Nrecload_FFT}. The active power deviation and ramp rate metrics are summarized in Table~\ref{Tab:ramp_rate_NREC_delay10}, and the band-by-band frequency fluctuation reduction results are reported in Table~\ref{Tab:band_analysis_NREC_10}. The results show that approximately 77.68\% of the fluctuation is concentrated in the 0.1-5 Hz band, 17.79\% is in the 5-10 Hz band, and 4.11\% is in the 10-30 Hz band. The proposed control effectively suppresses the fluctuations across the dominant frequency bands, achieving reductions of 94.1\%, 88.0\%, and 77.0\% in the 0.1-5 Hz, 5-10 Hz, and 10-30 Hz bands, respectively. The overall weighted broadband reduction is 93.1\%. In addition, the peak-to-peak active power excursion is reduced by 80.7\%, and the maximum ramp rate is reduced by 85.2\%, demonstrating the effectiveness of the proposed control.

\begin{table}
\vspace{-0.2cm}
\caption{Comparison of PCC active power deviations and ramp rates for the NREC AI training load profile.}
\vspace{-0.2cm}
\footnotesize
\setlength{\tabcolsep}{5pt}
\centering
\begin{tabular}{ c|c|c|c }
\hline
Metric & w/o BESS & with BESS & Reduction \\
\hline
Peak-to-peak excursion  & 166.39 MW   & 32.16 MW  & 80.7\% \\
99th-pct-based deviation  & 115.61 MW   & 11.41 MW  & 90.1\% \\
Max \(|dP/dt|\)  & 155.1 MW/s & 20.1 MW/s & 85.2\% \\
99th-pct \(|dP/dt|\)  & 147.9 MW/s & 7.1 MW/s & 95.2\% \\
\hline
\end{tabular}
\label{Tab:ramp_rate_NREC_delay10}
\end{table}

\begin{table}
\caption{Band-by-band analysis of PCC active power fluctuation reduction for the NREC AI training load profile.}
\vspace{-0.2cm}
\footnotesize
\setlength{\tabcolsep}{4.5pt}
\centering
\begin{tabular}{ c|c|c||c|c|c }
\hline
\shortstack{Frequency \\ band} & Weight & Reduction &
\shortstack{Frequency \\ band} & Weight & Reduction \\
\hline
0.1-5 Hz  & 77.68\% & 94.1\% & 5-10 Hz  & 17.79\% & 88.0\% \\
10-30 Hz  & 4.41\% & 77.0\% & 30-60 Hz  & 0.12\% & -167.0\% \\
\hline
\end{tabular}
\label{Tab:band_analysis_NREC_10}
\end{table}

\begin{figure}
    \centering
    \includegraphics[width=0.5\textwidth]{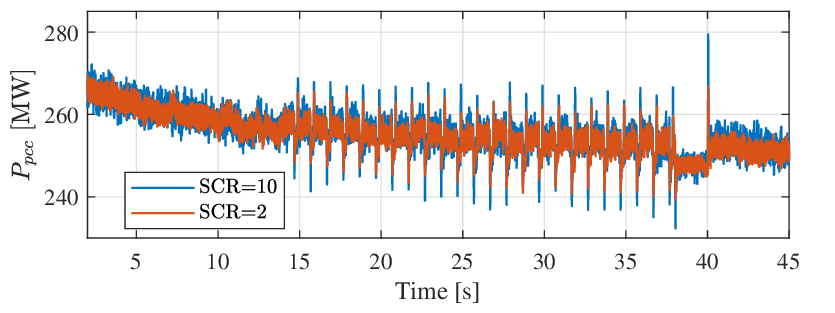}
    \vspace{-0.5cm}
    \caption{PCC active power for the BESS with the proposed control under different grid strengths.}
    \vspace{-0.3cm}
    \label{fig:different_SCR}
\end{figure}

To evaluate the impact of grid strength on the proposed control, simulations are conducted under different SCR conditions while the communication delay varies between 9 ms and 11 ms. Fig.~\ref{fig:different_SCR} shows the data center PCC active power for \(\mathrm{SCR}=10\) and \(\mathrm{SCR}=2\), with a 300 MW BESS equipped with the proposed control. The corresponding active power deviation and ramp rate reductions are summarized in Table~\ref{Tab:ramp_rate_ERCOT_different_SCR}, and the band-by-band frequency fluctuation reduction results are provided in Table~\ref{Tab:band_analysis_ERCOT_different_SCR}.
The results show that the proposed control maintains effective smoothing performance across different grid strength conditions. In particular, the weighted broadband reduction is 92.2\% for \(\mathrm{SCR}=10\) and 93.4\% \(\mathrm{SCR}=2\). The smoothing performance is slightly better under lower-SCR conditions. This is because, in a weaker grid, load-induced disturbances cause larger PCC voltage and frequency deviations, which naturally engage the GFM control more strongly. As a result, the BESS contributes more actively to power smoothing under weak grid conditions. 

\begin{table}
\caption{Comparison of PCC active power deviations and ramp rates with different SCR.}
\vspace{-0.2cm}
\centering
\begin{tabular}{ c|c|c  }
 \hline
Metric  & \shortstack{Reduction, SCR=2 } & \shortstack{Reduction, SCR=10  }\\
 \hline
Peak-to-peak excursion &  82.2\%   & 72.3\%\\
 99th-pct-based deviatio & 90.4\%  & 80.5\%\\
 Max $|dP/dt|$   & 87.3\% & 80.7\% \\
 99th-pct $|dP/dt|$  & 95.6\% & 94.6\% \\
 \hline
\end{tabular} \label{Tab:ramp_rate_ERCOT_different_SCR}
\vspace{-0.3cm}
\end{table}

\begin{table}
\caption{Band-by-band analysis of PCC active power fluctuation reduction with different SCR.}
\vspace{-0.2cm}
\centering
\begin{tabular}{ c|c|c}
 \hline
\shortstack {Frequency \\ band} & \shortstack{Reduction, SCR=2 } & \shortstack{Reduction, SCR=10  } \\
 \hline
 0.1-5 Hz  & 94.1\% & 93.8\%  \\
 5-10 Hz   & 88.7\% & 86.8\%  \\
 10-30 Hz  & 80.6\% & 63.9\% \\
 30-60 Hz  & -184.8\% & -425.6\% \\
 Weighted broadband &  93.4\%  & 92.2\%  \\
 \hline
\end{tabular} \label{Tab:band_analysis_ERCOT_different_SCR}
\vspace{-0.3cm}
\end{table}

\begin{figure}
    \centering
    \includegraphics[width=0.5\textwidth]{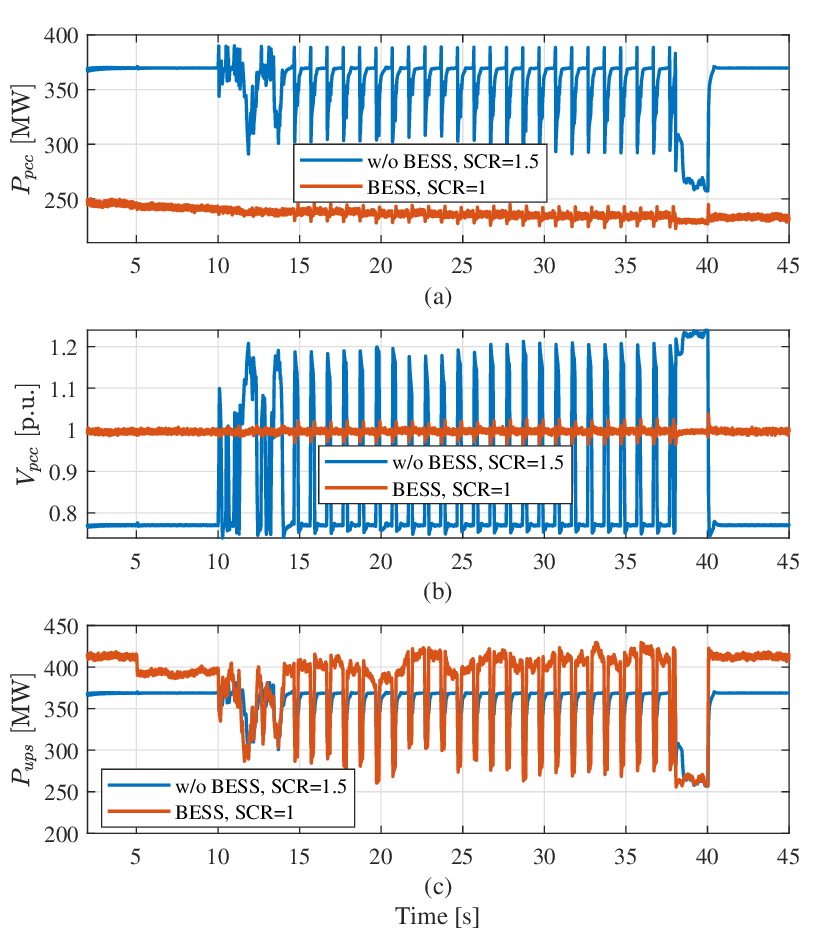}
    \vspace{-0.5cm}
    \caption{Simulation results under weak grid conditions without and with the BESS: (a) PCC active power, (b) PCC voltage magnitude, and (c) BESS active power.}
    \vspace{-0.3cm}
    \label{fig:SCR_1}
\end{figure}

Weak grid conditions with \(\mathrm{SCR}<2\) are also considered. Fig.~\ref{fig:SCR_1} presents the PCC active power, PCC voltage magnitude, and BESS active power for two weak grid cases: (1) \(\mathrm{SCR}=1.5\) without BESS and (2) \(\mathrm{SCR}=1\) with the BESS equipped with the proposed control. The results show that, when \(\mathrm{SCR}=1.5\) and no BESS is deployed, the rapid power fluctuations caused by AI training workloads can lead to significant voltage variations, indicating that the grid alone may not be able to adequately accommodate the highly dynamic power demand. In contrast, when \(\mathrm{SCR}=1\), the BESS with the proposed control maintains the data center PCC voltage through its GFM functionality. Consequently, the AI workload power demand is reliably supplied during different operating phases. The load current compensation and GFM control also work together effectively to significantly smooth the associated PCC power fluctuations.
\vspace{-0.2cm}
\subsection{Performance under Different AI Training Load Profiles}
\begin{figure}
    \centering
    \includegraphics[width=0.5\textwidth]{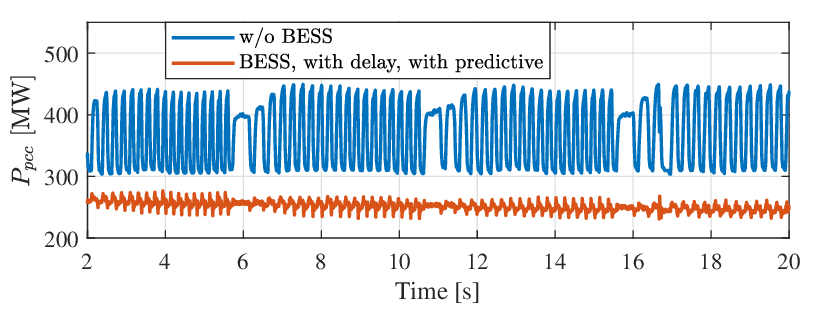}
    \vspace{-0.5cm}
    \caption{Data center PCC active power with and without the BESS for the high frequency AI training load profile.}
    \vspace{-0.3cm}
    \label{fig:NRL_data}
\end{figure}

\begin{table}
\caption{Comparison of PCC active power deviations and ramp rates for the high frequency AI training load profile.}
\vspace{-0.2cm}
\footnotesize
\setlength{\tabcolsep}{4.5pt}
\centering
\begin{tabular}{ c|c|c|c }
\hline
Metric & w/o BESS & with BESS & Reduction \\
\hline
Peak-to-peak excursion  & 145.79 MW   & 45.49 MW  & 68.8\% \\
99th-pct-based deviation  & 79.06 MW   & 18.50 MW  & 76.6\% \\
Max \(|dP/dt|\)  & 778.1 MW/s & 121.6 MW/s & 84.4\% \\
99th-pct \(|dP/dt|\)  & 671.2 MW/s & 86.3 MW/s & 87.1\% \\
\hline
\end{tabular}
\label{Tab:ramp_rate_modNREC_delay10}
\end{table}

\begin{figure}
    \centering
    \includegraphics[width=0.45\textwidth]{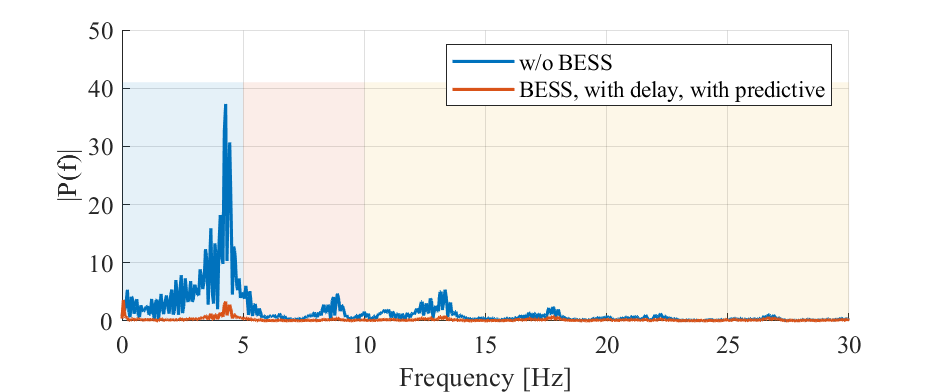}
    \vspace{-0.4cm}
    \caption{Single-sided amplitude spectrum of the PCC active power after DC component removal for the high frequency AI training load profile.}
    \vspace{-0.3cm}
    \label{fig:FFT_NREL}
\end{figure}

\begin{table}
\vspace{-0.2cm}
\caption{Band-by-band analysis of PCC active power fluctuation reduction for the high frequency AI training load profile.}
\vspace{-0.2cm}
\footnotesize
\setlength{\tabcolsep}{4.5pt}
\centering
\begin{tabular}{ c|c|c||c|c|c }
\hline
\shortstack{Frequency \\ band} & Weight & Reduction &
\shortstack{Frequency \\ band} & Weight & Reduction \\
\hline
0.1-5 Hz  & 75.45\% & 90.5\% & 5-10 Hz  & 15.04\% & 89.2\% \\
10-30 Hz  & 8.41\% & 77.0\% & 30-60 Hz  & 0.11\% & 8.4\% \\
\hline
\end{tabular}
\vspace{-0.3cm}
\label{Tab:band_analysis_modNREC_10}
\end{table}

To further validate the robustness of the proposed control, another representative AI training load profile is tested. 
A dataset of generative AI workload profiles is provided in~\cite{vercellino2026measurement}, with a sampling resolution of 0.2~s/0.1~s. One representative profile is selected and re-sampled at a higher resolution for EMT simulation. Under \(\mathrm{SCR}=3\) and a communication delay varying between 9 ms and 11 ms, the resulting data center PCC active power, with and without the BESS under the proposed control, is shown in Fig.~\ref{fig:NRL_data}. The quantitative metrics are summarized in Table~\ref{Tab:ramp_rate_modNREC_delay10}. The amplitude spectrum of the PCC active power is shown in Fig.~\ref{fig:FFT_NREL}. It can be observed that the dominant spectral component occurs around 5~Hz, indicating that this load profile contains relatively high frequency power fluctuations.
Considering the high frequency nature of this AI workload profile, a 200 ms time window is used for ramp-rate evaluation. The results show that the proposed control reduces the maximum ramp rate by 84.4\%, and the 99th-percentile ramp rate by 87.1\%. In terms of frequency domain performance, the overall weighted fluctuation reduction is 89.0\%, demonstrating the effectiveness of the proposed control for high frequency AI workload fluctuations.
\vspace{-0.3cm}
\section{Conclusion}
\vspace{-0.1cm}
This paper proposed a hybrid BESS control strategy to mitigate rapid power fluctuations caused by AI workloads. The proposed control combines instantaneous load current-based compensation with droop-based GFM control. The load-following controller provides fast power smoothing, while the GFM controller supports stable operation under different grid strength conditions and regulates the long term power output of the BESS.
The communication delay associated with transmitting load current measurements to the BESS controller was explicitly modeled, including time varying delay conditions. To mitigate the adverse impact of communication delay on smoothing performance, a predictor-based compensation method was developed and incorporated into the BESS control structure. Simulation studies were conducted under different communication delay ranges, grid strength conditions, and AI workload profiles. Both time domain and frequency domain analyses demonstrated that the proposed control effectively reduces data center PCC active power fluctuations and ramp rates while maintaining stable operation of the data center system. These results validate the effectiveness and robustness of the proposed communication delay robust control strategy for AI data center load smoothing.

\vspace{-0.3cm}
\section*{Acknowledgments}
\vspace{-0.1cm}
The authors would like to thank Sai Gopal Vennelaganti for providing valuable information on communication delays associated with BESS deployment in data centers.

\vspace{-0.2cm}
\bibliography{ref.bib}

\bibliographystyle{IEEEtran}

\vfill

\end{document}